\documentclass[a4paper]{article}

\usepackage{INTERSPEECH2022}

\usepackage{xcolor}
\usepackage{diagbox}
\usepackage{multirow}
\usepackage{booktabs}
\usepackage{subfigure}
\usepackage{mathtools}
\usepackage{amsmath}
\usepackage{cite}
\usepackage{hyperref}

\title{Disentanglement of Emotional Style and Speaker Identity for Expressive\\ Voice Conversion}
\name{Zongyang Du$^{1,2}$\thanks{\textbf{Speech Samples:} \href{https://zy-du.github.io/IS22/}{https://zy-du.github.io/IS22/}}, Berrak Sisman$^1$, Kun Zhou$^2$, Haizhou Li$^{3,2}$}
\address{$^1$Singapore University of Technology and Design, Singapore\\
  $^2$National University of Singapore, Singapore\\
  $^3$The Chinese University of Hong Kong, Shenzhen, China
  }
\email{zongyang\_du@mymail.sutd.edu.sg, berrak\_sisman@sutd.edu.sg, zhoukun@u.nus.edu, haizhouli@cuhk.edu.cn}

\begin{document}

\maketitle
\begin{abstract}
  Expressive voice conversion performs identity conversion for emotional speakers by jointly converting speaker identity and 
  emotional style. Due to the hierarchical structure of speech emotion, it is challenging to disentangle the 
  emotional style for different speakers. 
Inspired by the recent success of speaker disentanglement with variational autoencoder (VAE), we propose an any-to-any expressive voice conversion framework, that is called \textit{StyleVC}. StyleVC is designed to disentangle linguistic content, speaker identity, pitch, and emotional style information. We study the use of style encoder to model emotional style explicitly. 
At run-time, StyleVC converts both speaker identity and emotional style for arbitrary speakers. 
Experiments validate the effectiveness of our proposed framework in both objective and subjective evaluations.
\end{abstract}

\section{Introduction}
\begin{figure}[t]
    \centering
\vspace{-3mm}
    \subfigure[Traditional VC]{
    \begin{minipage}[t]{0.32\linewidth}
    \centering
    \includegraphics[width=2.5cm]{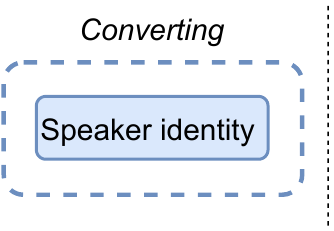}
    \label{VC}
    \end{minipage}%
    }%
    \subfigure[Emotional VC]{
    \begin{minipage}[t]{0.32\linewidth}
    \centering
    \includegraphics[width=2.5cm]{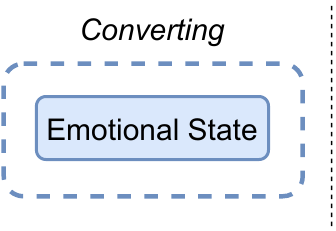}
    \label{EmoVC}
    \end{minipage}%
    }%
    \subfigure[\textbf{Expressive VC}]{
    \begin{minipage}[t]{0.32\linewidth}
    \centering
    \includegraphics[width=2.5cm]{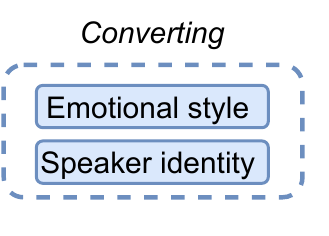}
    \label{ExpVC}
    \end{minipage}%
    }%
    \vspace{-3mm}
    \caption{A comparison of traditional VC, emotional VC and expressive VC at run-time. }
    \vspace{-4mm}
    \label{fig:diff}
\end{figure}

Human voice can be emotive and expressive. Emotion is expressed in speech to convey speaker motivation, mood and personality \cite{ming2016deep, shankar2019multi,9687906}. Expressive voice conversion is a task of jointly performing speaker identity and style transfer for emotional speakers while preserving the linguistic information, as illustrated in Figure \ref{ExpVC}.  

Voice conversion (VC) typically seeks to convert the speaker identity while preserving the linguistic content~\cite{sisman2020overview},  as illustrated in Figure \ref{VC}. Earlier studies include Gaussian mixture model (GMM) \cite{toda2007voice}, exemplar methods \cite{takashima2012exemplar} and sparse representation \cite{8688454}.
Deep learning methods, such as  deep neural network (DNN) \cite{xie2016kl} and recurrent neural network (RNN) \cite{nakashika2014high} have significantly improved the performance.
However, these frameworks require parallel training data, which limits their scope of applications. On the other hand, domain translation models, such as CycleGAN \cite{kaneko2018cyclegan} or StarGAN \cite{kameoka2018stargan} are studied for non-parallel training data. These models mostly rely on the cycle-consistency mechanism and they carry forward  the source speech style into the converted voices, thus not suitable for expressive voice conversion.

Another way of non-parallel VC is to learn the disentangled speech representations with VAE \cite{kingma2014auto}. VAE allows us to separately manipulate different disentangled features to achieve speaker identity conversion \cite{hsu2016voice,hsu2017voice,huang2018voice} or emotion conversion \cite{zhou2020converting,zhou2021vaw,zhou2021seen}. Other techniques to obtain a better disentangled representation include information bottleneck\cite{qian2019autovc,qian2020unsupervised}, instance normalization \cite{chou2019one,chen2021again}, and vector quantization (VQ) \cite{wu2020vqvc+, tjandra2019vqvae}. Recent studies, such as vector quantization and mutual information-based voice conversion (VQMIVC) \cite{wang2021vqmivc}, show the effectiveness of mutual information (MI) loss in reducing  the dependencies between different representations, which inspires this study.

A related technique is emotional voice conversion, which  converts the emotional state of the speaker while keeping the speaker identity unchanged \cite{zhou2021emotional, zhou2022emotion}, as illustrated in Figure \ref{EmoVC}. 
Unlike emotional voice conversion, expressive voice conversion seeks to jointly convert speaker identity and speech style for emotional speakers \cite{9687906}, as shown in Figure \ref{ExpVC}. 

As 
emotional style is an interplay among various speech elements~\cite{schuller2013computational,zhou2020converting}, such as speech content, speaker identity, speaking style and pitch, it is advantageous to disentangle the elements for effective expressive voice conversion. For example, such disentanglement allows us to carry over the speech content from the source to the target, but project the target speaker identity and speaking style of one certain emotional state into the target speech.  Since speech emotion is highly complex with multiple signal attributes concerning spectrum and prosody \cite{zhou2020converting}, it is challenging to model emotional style of arbitrary speakers. Most related studies, such as \cite{9687906, gan2022iqdubbing}, have only dealt with a multi-speaker expressive voice conversion model.

\begin{figure*}[t]
    \centering
    \includegraphics[width=18cm]{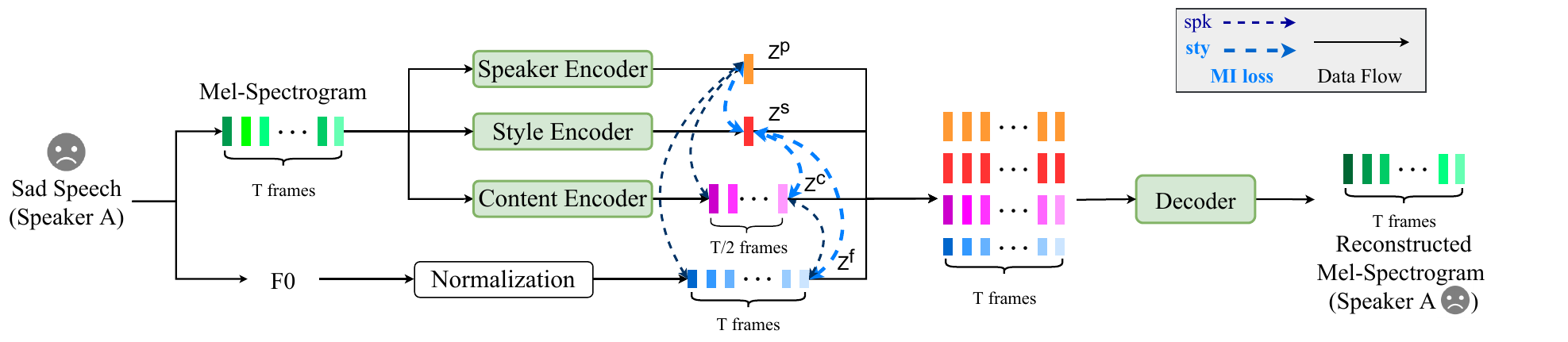}
    \vspace{-5mm}
    \caption{An illustration of the training phase of the proposed framework, where the green boxes represent the modules that are involved in the training. }
    \vspace{-4mm}
    \label{fig:train}
\end{figure*}

Our main contributions include: 1) we study the disentanglement of speaker identity and emotional style for expressive voice conversion; 2) we introduce a style encoder to explicitly model emotional style across different speakers, and further employ \textit{Style MI loss} to reduce the inter-dependency of emotional style and other representations (content, pitch, speaker); 3) the proposed method achieves  expressive voice conversion in an unsupervised way without any text transcriptions, speaker or emotion labels; 4) our proposed framework can effectively convert both speaker identity and emotional style  
 from arbitrary emotional speakers at run-time.  To our best knowledge, this is the first study for any-to-any expressive voice conversion.

The rest of this paper is organized as follows: 
Section 2 presents the details of our proposed framework. 
In Section 3, we report the experimental results. Section 4 concludes the study.

\section{Method}

We propose an any-to-any expressive voice conversion framework, \textit{StyleVC}, which can jointly perform both emotional style and speaker identity conversion. VQMIVC \cite{wang2021vqmivc} is a voice conversion framework based on VAE \cite{kingma2014auto} performing an unsupervised speaker disentanglement. We extend the idea of VQMIVC from speaker identity disentanglement to emotional style disentanglement. 


StyleVC consists of a style encoder, a speaker encoder, a content encoder and a decoder as shown in Figure \ref{fig:train}.
During training, StyleVC learns to disentangle emotional style by employing an emotional style encoder to explicitly extract style representation and adopting mutual information as the correlation metric to reduce the dependencies between emotional style representation and other representations. At run-time, StyleVC allows us to convert both speaker identity and emotional style for any speaker by manipulating different disentangled speech representations. 




\vspace{-2mm}
\subsection{Training Phase}
Given an expressive utterance, we first extract Mel-spectrograms $X = \{x_1,x_2,...x_T\}$ and fundamental frequency $F_{0}$ where $T$ is the total number of frames. The style encoder $E^{s}$ learns to encode the Mel-spectrograms into a fixed-length representation $Z^s$ = $E^{s}(X)$. The $Z^s$ represents the emotional style in the utterance level. The content encoder $E^c$ extracts content $Z^c$ = $\{z_1^c,z_2^c, ...,z_{T/2}^c\}$ from $X$. The speaker encoder $E^{p}$ learns to embed the Mel-spectrograms into a fixed-length speaker embedding: $Z^p$ = $E^{p}(X)$. To represent the intonation, F0 is extracted from the speech waveform and log normalized into zero mean and unit variance. Since the F0 varies with the speakers, we take the log normalized F0 as the pitch embedding $Z^f$, and study it separately.

We note that the speaker embedding $Z^p$ and the emotional style embedding $Z^s$ represent the speaker identity and the emotional style information at an utterance level. To align with the pitch embedding $Z^f$, we up-sample speaker embedding $Z^p$, the emotional style embedding $Z^s$, and the content embedding $Z^c$ to $T$ frames. The decoder $D$ aims to reconstruct acoustic features $\hat{X}$ from pitch embedding $Z^f$ and the upsampled speech embedding. A reconstruction loss is calculated between the reconstructed Mel-spectrogram and the ground-truth. 

During training, the correlation among different speech representations can be reduced by minimizing the MI loss as follows: 

1) \textbf{Speaker MI Loss}: A speaker MI loss is applied to speaker style embedding $Z^p$, content embedding $Z^c$ and pitch embedding $Z^f$ as follows:
\begin{equation}
\begin{aligned} 
\vspace{-3mm}
L_{spk-MI} = \hat{I}(Z^p,Z^c)+ \hat{I}(Z^p,Z^f)+\hat{I}(Z^c,Z^f)
\end{aligned}
\label{SPKMI}
\end{equation}
where $\hat{I}$ represents the unbiased estimation for vCLUB as described in \cite{wang2021vqmivc}. Speaker MI loss is effective for alleviating the leakage of content and pitch information into the speaker representation from \cite{wang2021vqmivc}. Motivated by this, we incorporate style mutual information minimization to disentangle emotional style representation and other representations. 

2) \textbf{Style MI Loss:} A style MI loss is applied to emotional style embedding $Z^s$, speaker embedding $Z^p$, content embedding $Z^c$ and pitch embedding $Z^f$ as follows:
\begin{equation}
\begin{aligned} 
\vspace{-3mm}
L_{sty-MI} &= \hat{I}(Z^s,Z^p)+ \hat{I}(Z^s,Z^c)+ \hat{I}(Z^s,Z^f)
\end{aligned}
\label{STYMI}
\end{equation}

The total MI loss is given as follows:

\begin{equation}
\begin{aligned} 
\vspace{-3mm}
L_{MI} =\lambda_{sty}L_{sty-MI}+\lambda_{spk}L_{spk-MI}
\end{aligned}
\label{loss}
\end{equation}
where $\lambda_{sty}$, and $\lambda_{spk}$ are the weights to control how MI loss enhances the disentanglement. We believe emotional style representation carries more correlation information than other representations. {We set  $\lambda_{sty}$ = 2$\lambda_{spk}$ to reduce the correlation information between emotional style representation and other representations.}

\begin{figure*}
    \centering
    \subfigure[Neutral]{
    \begin{minipage}[t]{0.2\linewidth}
    \centering
    \includegraphics[width=3.7cm]{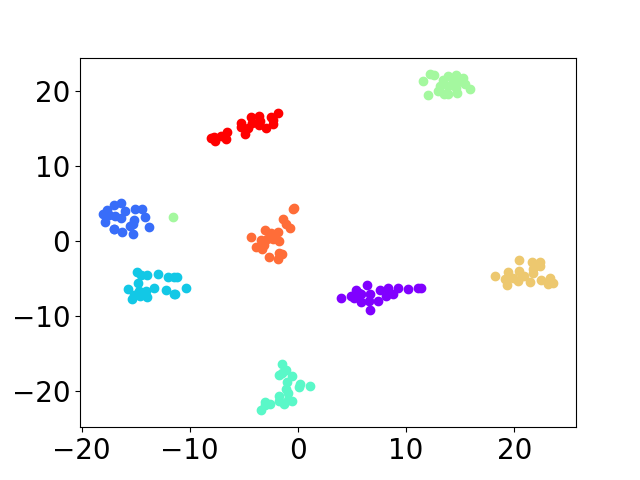}
    \end{minipage}%
    }%
    \subfigure[Happy]{
    \begin{minipage}[t]{0.2\linewidth}
    \centering
    \includegraphics[width=3.7cm]{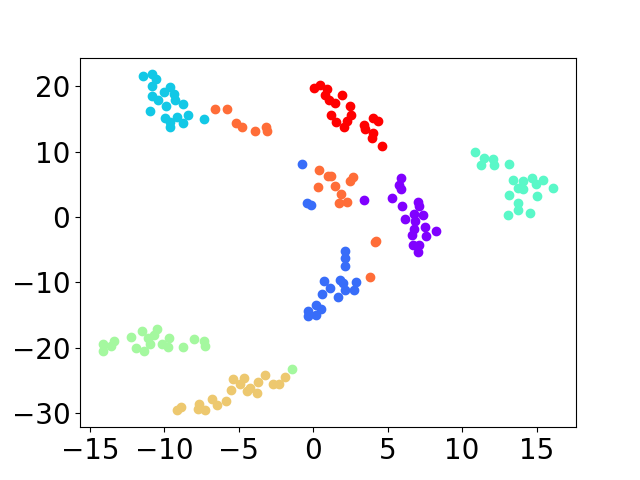}
    \end{minipage}%
    }%
    \subfigure[Angry]{
    \begin{minipage}[t]{0.2\linewidth}
    \centering
    \includegraphics[width=3.7cm]{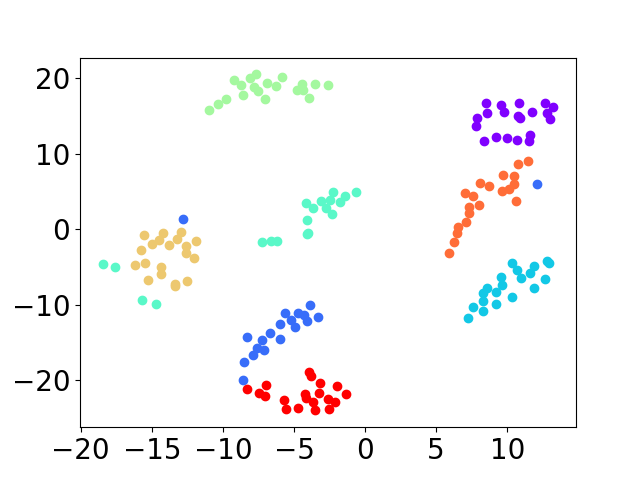}
    \end{minipage}
    }%
    \subfigure[Surprise]{
    \begin{minipage}[t]{0.2\linewidth}
    \centering
    \includegraphics[width=3.7cm]{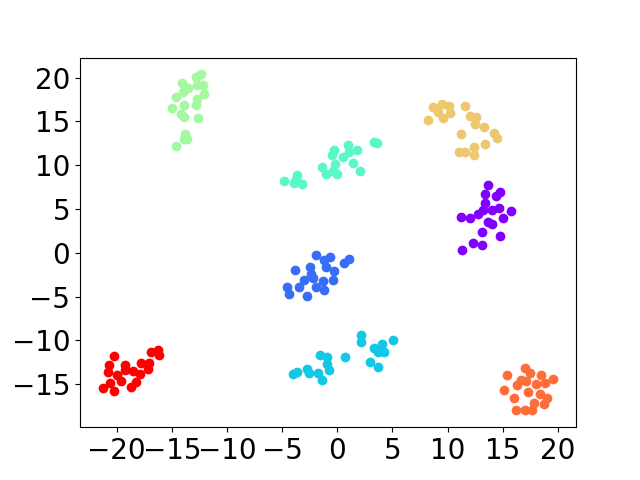}
    \end{minipage}
    }%
    \subfigure[Sad]{
    \begin{minipage}[t]{0.2\linewidth}
    \centering
    \includegraphics[width=3.7cm]{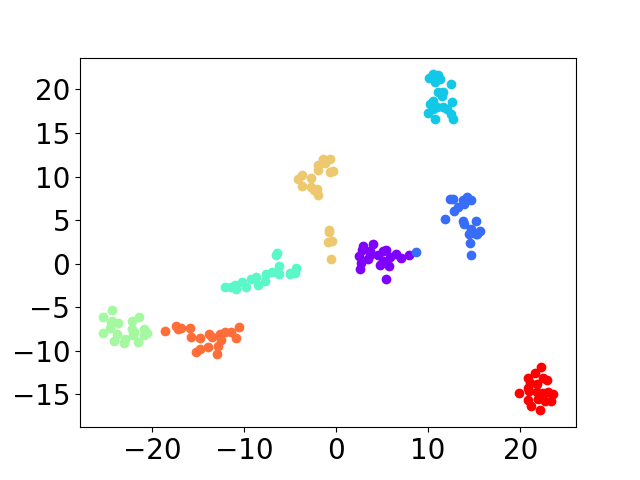}
    \end{minipage}
    }%
    \centering
    \caption{Visualization of speaker identity representations of 8 seen speakers for 5 different emotions. Each color represents a different speaker.}
    \vspace{-4mm}
\label{fig:visualization}
\end{figure*}

\begin{figure}[t]
    \centering
    \includegraphics[width=8cm]{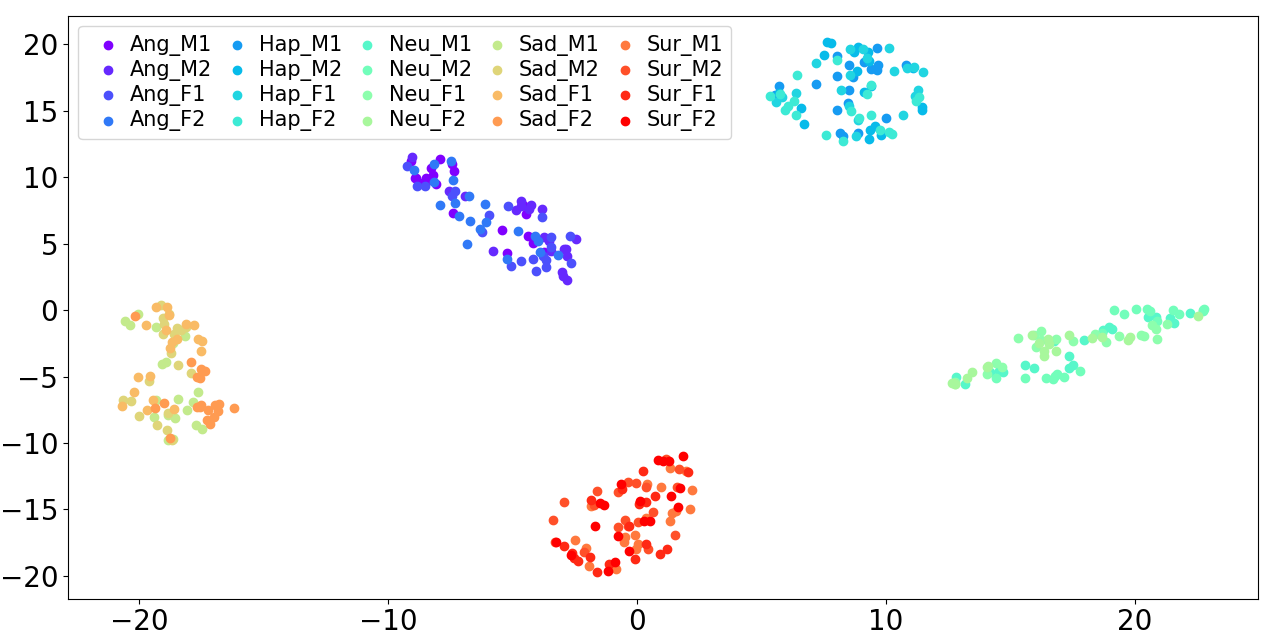}
    \vspace{-5mm}
    \caption{Visualization of emotional style representations of 4 seen speakers speaking with 5 different emotions.}
    \vspace{-4mm}
    \label{fig:emo_emb}
\end{figure}

\subsection{Run-time Conversion}

{At run-time, StyleVC takes a source utterance from one speaker as input, and a reference utterance from another speaker, who is either seen or unseen during training. The content encoder generates the source content embedding from the source utterance. Given the reference emotional utterance from the target speaker, we use the speaker encoder and style encoder to generate speaker and emotional style embedding respectively. 
We then convert the F0 with the mean and standard variance calculated from a random seen male/female speaker. We expect that the emotional style embedding can capture emotional style that is related to the target speaker.
The decoder learns to generates the converted Mel-spectrograms with the source content embedding, the speaker and emotional style embedding from one target speaker's utterance, and the converted F0.  The speech waveform is reconstructed with Parallel WaveGAN vocoder \cite{yamamoto2020parallel}.}

\subsection{Emotional Style and Speaker Disentanglement}
We analyze the effectiveness of StyleVC on both emotional style disentanglement and speaker identity disentanglement by visualizing the generated speaker embedding  from speaker encoder with t-SNE \cite{van2008visualizing} in Figure \ref{fig:visualization}.
 {We perform the experiments on ESD database \cite{zhou2021seen}, and use 20 utterances for each seen speaker and each emotion. }
We observe that each speaker forms an identifiable cluster for each emotion that suggests that we obtain effective speaker embedding. 

 We further visualize the emotional style features of 4 seen speakers from 5 different emotions. 
 {For each speaker and each emotion, we use 20 utterances.}
 Figure \ref{fig:emo_emb} shows that utterances in different emotions are well-separated and utterances spoken by different speakers in the same emotion are mixed-up in every cluster, that suggests we obtain effective speaker-independent emotional style embeddings.  


\section{Experiments}

\begin{table*}[!t]
\vspace{-5mm}
\centering
\caption{Objective evaluation results for S2S, S2U and U2U settings}
\vspace{-2mm}
\scalebox{0.85}{
\begin{tabular}{cc|cc|ccc|cc}
\hline
                                                       &          & \multicolumn{2}{c|}{MCD{[}dB{]}} & \multicolumn{3}{c|}{SV accuracy {[}\%{]}}     & \multicolumn{2}{c}{F0 -RMSE {[Hz]}} \\
                                                       &          & VQMIVC \cite{wang2021vqmivc}          & StyleVC        & VQMIVC \cite{wang2021vqmivc}           & StyleVC          & Target  & VQMIVC\cite{wang2021vqmivc}          & StyleVC          \\ \hline
\multicolumn{1}{c|}{\multirow{5}{*}{Seen to Seen (S2S)}}     & Happy    & 6.28           & 6.08 & 85.19         & 86.62 & 95.70 & 42.43          & 39.38  \\
\multicolumn{1}{c|}{}                                  & Sad      & 5.83           & 5.69 & 85.41          & 89.17 & 94.46 & 43.39          & 40.81  \\
\multicolumn{1}{c|}{}                                  & Angry    & 6.98           & 6.77 & 86.18          & 89.77 & 93.17 & 54.36          & 49.66  \\
\multicolumn{1}{c|}{}                                  & Surprise & 6.61           & 6.57 & 81.47          & 86.44 & 93.73 & 47.83          & 43.51  \\
\multicolumn{1}{c|}{}                                  & Neutral  & 5.82           & 5.79 & 88.41          & 90.82 & 94.98 & 36.31          & 34.69  \\ \hline
\multicolumn{1}{c|}{\multirow{5}{*}{Seen to Unseen (S2U)}}   & Happy    & 6.49  & 6.67          & 73.41          & 77.68 & 93.12 & 73.94          & 71.99  \\
\multicolumn{1}{c|}{}                                  & Sad      & 7.45           & 7.19 & 66.72 & 65.86          & 91.85 & 34.27          & 30.39  \\
\multicolumn{1}{c|}{}                                  & Angry    & 7.57           & {7.33} & 76.43          & {80.82} & 91.94 & 36.67          & {36.07}  \\
\multicolumn{1}{c|}{}                                  & Surprise & 7.63           & {7.47} & 78.46          & {83.69} & 93.86 & 76.02          & {68.41}  \\
\multicolumn{1}{c|}{}                                  & Neutral  & 7.51           & {7.36} & 71.54          & {78.85} & 94.15 & 66.45          & {62.96}  \\ \hline
\multicolumn{1}{c|}{\multirow{5}{*}{Unseen to Unseen (U2U)}} & Happy    & 7.49           & {7.37} & 75.06          & {75.45} & 95.03 & 69.99          & {63.23}  \\
\multicolumn{1}{c|}{}                                  & Sad      & 6.82           & {6.82} & 64.84          & {66.84} & 95.03 & 55.54          & {55.17}  \\
\multicolumn{1}{c|}{}                                  & Angry    & 7.68           & {7.52} & 72.99          & {79.03} & 96.52 & 61.32          & {63.40} \\
\multicolumn{1}{c|}{}                                  & Surprise & {8.09}  & 8.32          & 74.98          & {80.23} & 91.58 & {86.85} & 82.32           \\
\multicolumn{1}{c|}{}                                  & Neutral  & 6.81           & {6.79} & 67.63          & {71.19} & 95.14 & 42.49          & {40.78} \\ \hline
\end{tabular}%
}
\vspace{-4mm}
\label{tab: objective}
\end{table*}
We conduct both objective and subjective evaluations on the ESD dataset\cite{zhou2021emotional} to assess the performance the speaker identity and emotional style conversion. {We use the speech data of 8  speakers from 5 emotions (neutral, happy, sad, angry and surprise) to train the networks. We choose the other 2 speakers as unseen speakers and conduct conversion on both seen and unseen speakers to assess the performance of our proposed model.} 
For each seen speaker from each emotion, we use 300 utterances for training, 25 utterances for validation. For evaluation, we use 25 utterances from each seen and unseen speakers. In a comparative study, we adopt VQMIVC \cite{wang2021vqmivc} as our baseline. 

\vspace{-2mm}
\subsection{Experimental Setup}
All the speech data is sampled at 16 kHz and saved in 16 bits. We extract 80-dimensional Mel-spectrograms and one-dimensional F0 as the acoustic features.
At run-time, F0 is converted through the logarithm Gaussian (LG) normalized transformation \cite{toda2007voice}.
The style encoder consists of a 6-layer stack of 2D convolutions with batch normalization (BN) and ReLu activation, a GRU layer, and two fully connected (FC) layers followed by ReLU. The content encoder contains a CNN layer with the stride of 2, 512-dimensional LC layer, a codebook with 512 64-dimensional learnable vectors and a 256-dimensional RNN layer. The speaker encoder consists of 8 ConvBank layers, 12 CNN layers and 4 linear layers. The decoder has an LSTM layer with 1024 nodes, 3 CNN layers, 2 1024-dimensional LSTM layers and an 80-dimensional linear layer. The whole framework is optimized with Adam with 15-epoch warm-up. We set the learning rate to 1e-3, and half it every 100 epochs. The total number of epochs is 800 with a batch size of 128. 
We use a publicly available version\footnote{https://github.com/kan-bayashi/ParallelWaveGAN} of Parallel WaveGAN as the vocoder, and train it with ESD dataset.
We set $\lambda_{spk}=0.01$ and $\lambda_{sty}=0.02$.

We evaluate our model in three scenarios: the conversion between seen speakers, denoted as \textit{S2S}; the conversion between seen speakers and unseen speakers, denoted as \textit{S2U}; and the conversion between unseen speakers, denoted as \textit{U2U}.

\subsection{Objective Evaluation}
We calculate Mel-cepstral distortion (MCD) to measure the spectral distortion in Table \ref{tab: objective}. We observe that our proposed framework consistently outperforms the baseline by achieving lower MCD values for S2S, S2U and U2U. This indicates the effectiveness of our proposed framework. 

We then calculate F0-RMSE to evaluate 
pitch disentanglement performance. From Table \ref{tab: objective}, we observe that our proposed framework always achieves a better performance than the baseline for all the emotions
by obtaining a lower F0-RMSE values. This observation validates the effectiveness of our proposed framework in terms of the prosody conversion. 

An open-source pre-trained speaker verification model\footnote{https://github.com/resemble-ai/Resemblyzer} is utilized to measure the speaker similarity between a converted utterance and a target utterance, as conducted in previous studies \cite{gan2022iqdubbing, Lin2021S2VCAF}. We observe that StyleVC outperforms the baseline in terms of speaker identity disentanglement by achieving higher SV accuracy. These results indicate the superior performance of our proposed framework in terms of the speaker identity conversion. 

\begin{table}[t]
\centering
    \caption{MOS scores for speech quality by 14 listeners.}
    \vspace{-2mm}
    \scalebox{0.85}{
    \begin{tabular}{c|ccc}
\hline
\multirow{2}{*}{Framework} & \multicolumn{3}{c}{MOS}                                                               \\ \cline{2-4} 
                           & \multicolumn{1}{c|}{S2S}     & \multicolumn{1}{c|}{S2U} & U2U \\ \hline
VQMIVC                     & \multicolumn{1}{c|}{$3.45\pm0.13$}          & \multicolumn{1}{c|}{$3.42\pm0.13$}            & {$3.63\pm0.15$}               \\ 
StyleVC            & \multicolumn{1}{c|}{3.54 $\pm$ 0.14} & \multicolumn{1}{c|}{3.58 $\pm$ 0.15}            &   {3.74 $\pm$ 0.15}             \\ \hline
\end{tabular}}
    \label{tab:MOS}
    \vspace{-3mm}
\end{table}

\begin{table}[t]
\centering
\caption{MCD [dB] values for seen speakers with angry emotion. }
\vspace{-3mm}
\label{tab:ab}
\scalebox{0.9}{
\begin{tabular}{c|cc|c}
\hline
           & \begin{tabular}[c]{@{}c@{}}Intra-gender\end{tabular} & \begin{tabular}[c]{@{}c@{}}Inter-gender\end{tabular} & Average        \\ \hline
w/o  & 6.34                                                   & 6.98                                                   & 6.66          \\ 
+$I_{Z^s,Z^p}$       & 6.32                                                   & 6.94                                                   & 6.63          \\ 
+$I_{Z^s,Z^c}$       & 6.38                                                   & 6.74                                                   & 6.56          \\ 
+$I_{Z^s,Z^f}$       & 6.45                                                   & 6.78                                                   & 6.62          \\ 
+$I_{Z^s,Z^c}$ +$I_{Z^s,Z^f}$    & 6.40                                                   & 6.83                                                   & 6.62          \\ 
+$I_{Z^s,Z^p}$+$I_{Z^s,Z^c}$     & 6.53                                                   & 6.86                                                   & 6.70          \\
+$I_{Z^s,Z^p}$+$I_{Z^s,Z^f}$    & 6.54                                                   & 7.44                                                   & 6.99          \\ \hline
+$L_{sty-MI}$ & 6.33                                                   &6.77                                                 & 6.55 \\ \hline
\end{tabular}%
}
\vspace{-4mm}
\end{table}

\vspace{-2mm}
\subsection{Subjective Evaluation}
We conduct listening experiments to assess the speech quality, speaker similarity and emotional style similarity. 14 subjects participate in all the experiments.  Each of them listens to 162 converted utterances in total.  

We first report the mean opinion scores (MOS) for S2S, S2U and U2U in Table \ref{tab:MOS}, where a higher score indicates a better speech quality. As shown in Table \ref{tab:MOS}, the StyleVC framework outperforms the baseline in terms of speech quality, which validates the proposed speaker identity conversion and emotional style conversion.

\begin{figure}[t]
    \centering
    \includegraphics[width=6.8cm]{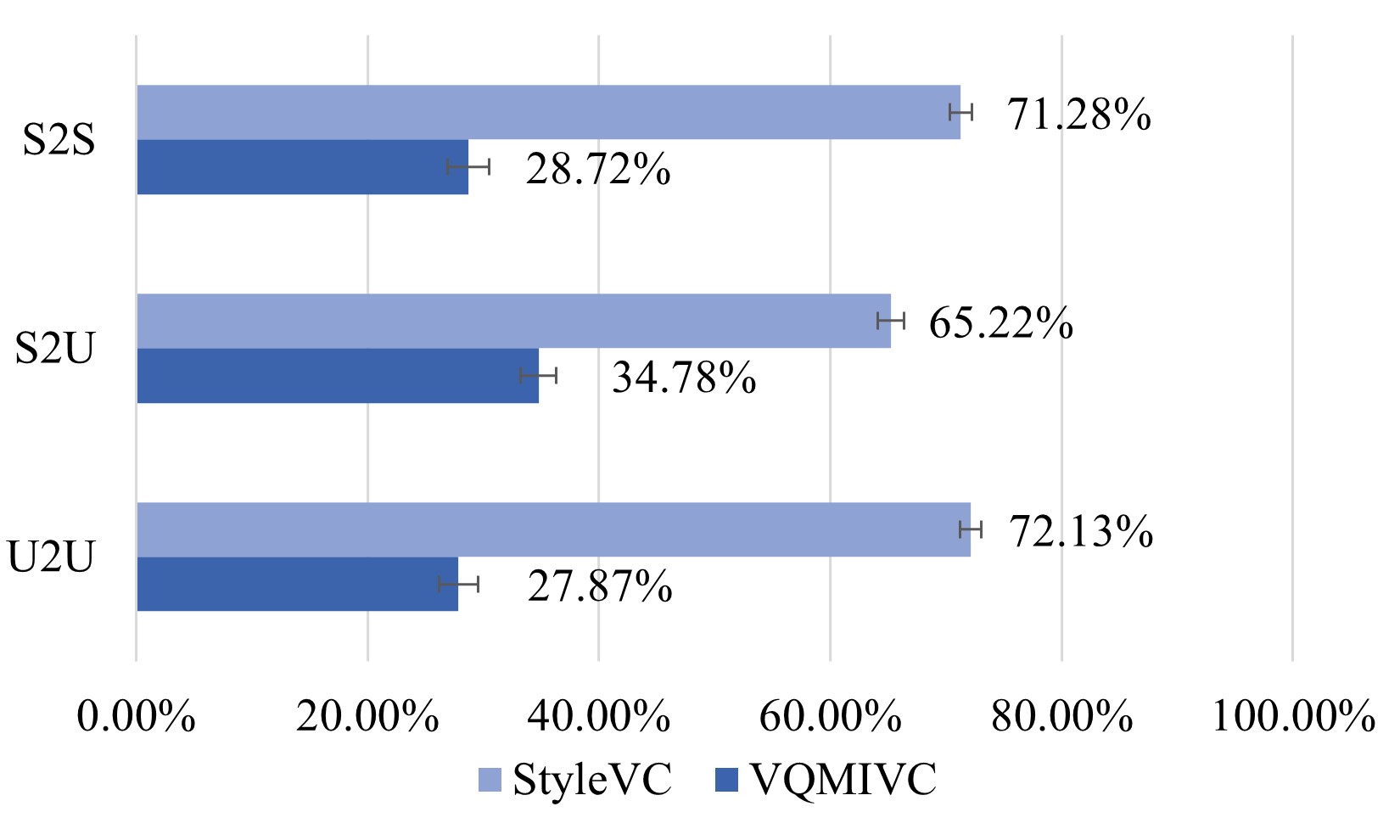}
    \vspace{-4mm}
    \caption{ABX preference results for speaker similarity with 95\% confidence interval.}
    \label{fig:xab1}
    \vspace{-5mm}
\end{figure}

We then conduct three ABX preference tests to evaluate the speaker similarity in S2S, S2U and U2U settings, where all the listeners are asked to listen to the reference and the converted utterances respectively, and choose the one that sounds closer to the reference in terms of the speaker identity. As shown in Figure \ref{fig:xab1}, our proposed framework outperforms the baseline, which suggests our proposed method has a much better performance on the speaker identity conversion than baseline. 

\begin{figure}
    \centering
    \includegraphics[width=6.8cm]{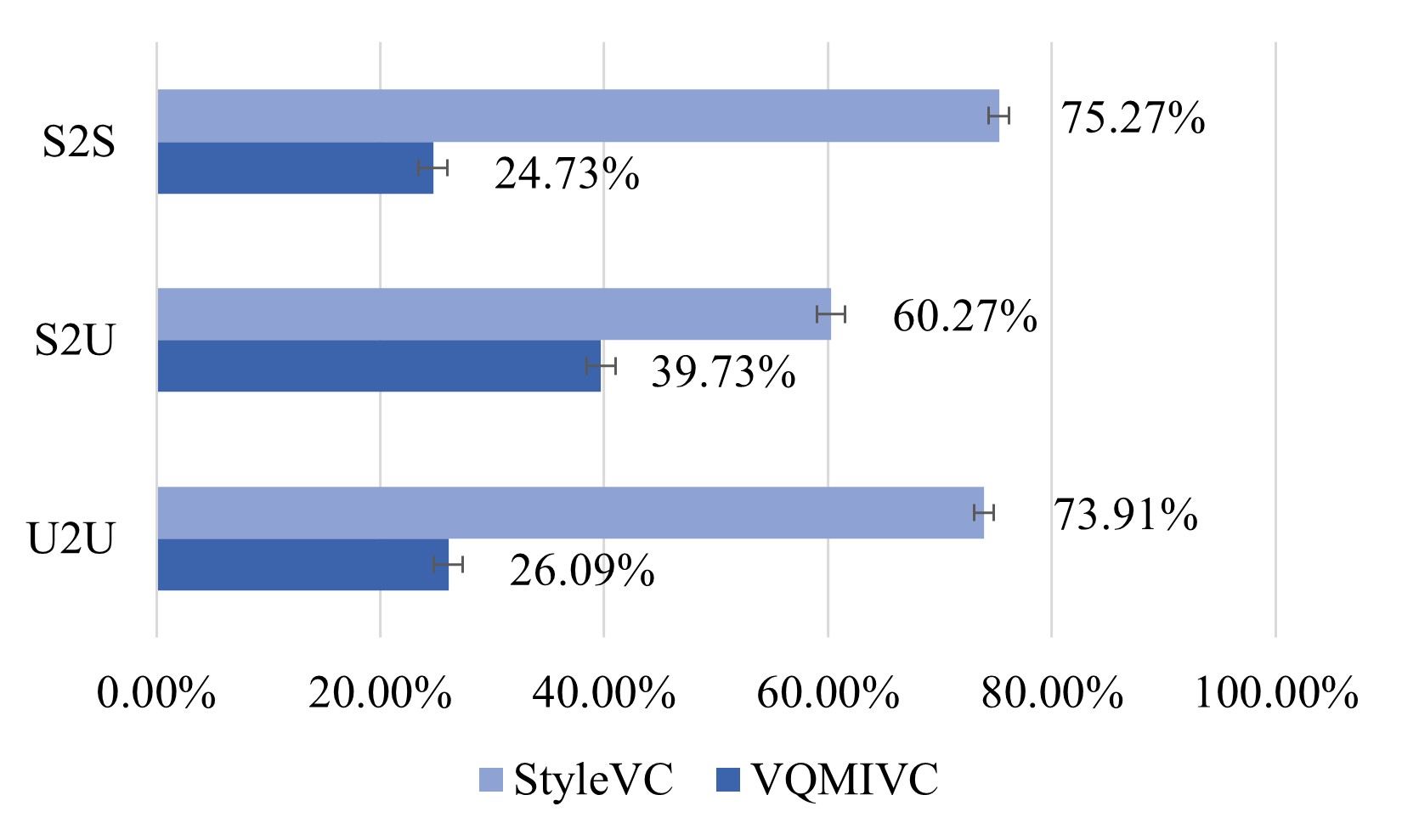}
       \vspace{-3mm}
    \caption{ABX preference results for emotional style similarity with 95\% confidence interval.}
    \label{fig:xab2}
    \vspace{-5mm}
\end{figure}

We further conduct three ABX preference tests to evaluate the emotional style similarity in S2S, S2U and U2U settings. In Figure \ref{fig:xab2}, we observe that our proposed framework significantly outperforms the baseline in terms of emotional style similarity. This suggests that our proposed framework is capable of converting emotional style for arbitrary speakers, which further validates our idea on emotional style conversion.
\vspace{-1mm}

\vspace{-2mm}
\subsection{Ablation studies}
We conduct ablation studies to validate the effectiveness of each term in Style MI loss, as shown in eq (\ref{STYMI}). We calculate MCD on converted speech of seen speakers from angry emotion. We report the results in Table \ref{tab:ab}. From the results, we observe Style MI loss, the combination of three MI terms ($I_{Z^s,Z^p}$, $I_{Z^s,Z^c}$, $I_{Z^s,Z^f}$) outperforms the baseline with only \textit{Speaker MI loss} in both inter-gender and intra-gender settings. And it achieves best performance among other combinations of MI terms with lower MCD values in the average setting.
\section{Conclusion}
In this paper, we study the disentanglement of speaker identity and emotional speech style for any-to-any expressive voice conversion. We propose a framework named \textit{StyleVC} to jointly convert the speaker identity and emotional style. We introduce a style encoder to explicitly model emotional style, and use MI losses to eliminate the shared information between different speech representations. Experimental results show that our proposed framework outperforms the baseline. Future directions include the study of duration modeling for expressive voice conversion.

\vspace{-2mm}
\section{Acknowledgment}
This research is supported by the Ministry of Education, Singapore, under its MOE Tier 2 funding programme, award no: MOE-T2EP50220-0021, SUTD Startup Grant Artificial Intelligence for Human Voice Conversion (SRG ISTD 2020 158) and SUTD AI Grant – Thrust 2 Discovery by AI (SGPAIRS1821). The research by Haizhou Li and Kun Zhou is supported by the Science and Engineering Research Council, Agency for Science, Technology and Research (A*STAR), Singapore, through the National Robotics Program under Human-Robot Interaction Phase 1 (Grant No. 192 25 00054), Human Robot Collaborative AI under its AME Programmatic Funding Scheme (Project No. A18A2b0046); the Guangdong Provincial Key Laboratory of Big Data Computing under the Grant No. B10120210117-KP02, The Chinese University of Hong Kong, Shenzhen (CUHK-Shenzhen). 

\bibliographystyle{IEEEtran}

\bibliography{references}

\begin{thebibliography}{10}
\providecommand{\url}[1]{#1}
\csname url@samestyle\endcsname
\providecommand{\newblock}{\relax}
\providecommand{\bibinfo}[2]{#2}
\providecommand{\BIBentrySTDinterwordspacing}{\spaceskip=0pt\relax}
\providecommand{\BIBentryALTinterwordstretchfactor}{4}
\providecommand{\BIBentryALTinterwordspacing}{\spaceskip=\fontdimen2\font plus
\BIBentryALTinterwordstretchfactor\fontdimen3\font minus
  \fontdimen4\font\relax}
\providecommand{\BIBforeignlanguage}[2]{{%
\expandafter\ifx\csname l@#1\endcsname\relax
\typeout{** WARNING: IEEEtran.bst: No hyphenation pattern has been}%
\typeout{** loaded for the language `#1'. Using the pattern for}%
\typeout{** the default language instead.}%
\else
\language=\csname l@#1\endcsname
\fi
#2}}
\providecommand{\BIBdecl}{\relax}
\BIBdecl

\bibitem{ming2016deep}
H.~Ming, D.~Huang, L.~Xie, J.~Wu, M.~Dong, and H.~Li, ``Deep bidirectional lstm
  modeling of timbre and prosody for emotional voice conversion,'' 2016.

\bibitem{shankar2019multi}
R.~Shankar, J.~Sager, and A.~Venkataraman, ``A multi-speaker emotion morphing
  model using highway networks and maximum likelihood objective.'' in
  \emph{INTERSPEECH}, 2019, pp. 2848--2852.

\bibitem{9687906}
Z.~Du, B.~Sisman, K.~Zhou, and H.~Li, ``Expressive voice conversion: A joint
  framework for speaker identity and emotional style transfer,'' in \emph{2021
  IEEE Automatic Speech Recognition and Understanding Workshop (ASRU)}, 2021,
  pp. 594--601.

\bibitem{sisman2020overview}
B.~Sisman, J.~Yamagishi, S.~King, and H.~Li, ``An overview of voice conversion
  and its challenges: From statistical modeling to deep learning,''
  \emph{IEEE/ACM Transactions on Audio, Speech, and Language Processing}, 2020.

\bibitem{toda2007voice}
T.~Toda, A.~W. Black, and K.~Tokuda, ``Voice conversion based on
  maximum-likelihood estimation of spectral parameter trajectory,'' \emph{IEEE
  Transactions on Audio, Speech, and Language Processing}, vol.~15, no.~8, pp.
  2222--2235, 2007.

\bibitem{takashima2012exemplar}
R.~Takashima, T.~Takiguchi, and Y.~Ariki, ``Exemplar-based voice conversion in
  noisy environment,'' in \emph{2012 IEEE Spoken Language Technology Workshop
  (SLT)}.\hskip 1em plus 0.5em minus 0.4em\relax IEEE, 2012, pp. 313--317.

\bibitem{8688454}
B.~Sisman, M.~Zhang, and H.~Li, ``Group sparse representation with wavenet
  vocoder adaptation for spectrum and prosody conversion,'' \emph{IEEE/ACM
  Transactions on Audio, Speech, and Language Processing}, vol.~27, no.~6, pp.
  1085--1097, 2019.

\bibitem{xie2016kl}
F.-L. Xie, F.~K. Soong, and H.~Li, ``A kl divergence and dnn-based approach to
  voice conversion without parallel training sentences.'' in
  \emph{Interspeech}, 2016, pp. 287--291.

\bibitem{nakashika2014high}
T.~Nakashika, T.~Takiguchi, and Y.~Ariki, ``High-order sequence modeling using
  speaker-dependent recurrent temporal restricted boltzmann machines for voice
  conversion,'' in \emph{Fifteenth annual conference of the international
  speech communication association}, 2014.

\bibitem{kaneko2018cyclegan}
T.~Kaneko and H.~Kameoka, ``Cyclegan-vc: Non-parallel voice conversion using
  cycle-consistent adversarial networks,'' in \emph{2018 26th European Signal
  Processing Conference (EUSIPCO)}.\hskip 1em plus 0.5em minus 0.4em\relax
  IEEE, 2018, pp. 2100--2104.

\bibitem{kameoka2018stargan}
H.~Kameoka, T.~Kaneko, K.~Tanaka, and N.~Hojo, ``Stargan-vc: Non-parallel
  many-to-many voice conversion using star generative adversarial networks,''
  in \emph{2018 IEEE Spoken Language Technology Workshop (SLT)}.\hskip 1em plus
  0.5em minus 0.4em\relax IEEE, 2018, pp. 266--273.

\bibitem{kingma2014auto}
D.~P. Kingma and M.~Welling, ``Auto-encoding variational bayes,'' \emph{stat},
  vol. 1050, p.~1, 2014.

\bibitem{hsu2016voice}
C.-C. Hsu, H.-T. Hwang, Y.-C. Wu, Y.~Tsao, and H.-M. Wang, ``Voice conversion
  from non-parallel corpora using variational auto-encoder,'' in \emph{2016
  Asia-Pacific Signal and Information Processing Association Annual Summit and
  Conference (APSIPA)}.\hskip 1em plus 0.5em minus 0.4em\relax IEEE, 2016, pp.
  1--6.

\bibitem{hsu2017voice}
------, ``Voice conversion from unaligned corpora using variational
  autoencoding wasserstein generative adversarial networks,'' \emph{arXiv
  preprint arXiv:1704.00849}, 2017.

\bibitem{huang2018voice}
W.-C. Huang, H.-T. Hwang, Y.-H. Peng, Y.~Tsao, and H.-M. Wang, ``Voice
  conversion based on cross-domain features using variational auto encoders,''
  in \emph{2018 11th International Symposium on Chinese Spoken Language
  Processing (ISCSLP)}.\hskip 1em plus 0.5em minus 0.4em\relax IEEE, 2018, pp.
  51--55.

\bibitem{zhou2020converting}
K.~Zhou, B.~Sisman, M.~Zhang, and H.~Li, ``Converting anyone’s emotion:
  Towards speaker-independent emotional voice conversion,'' \emph{Proc.
  Interspeech 2020}, pp. 3416--3420, 2020.

\bibitem{zhou2021vaw}
K.~Zhou, B.~Sisman, and H.~Li, ``Vaw-gan for disentanglement and recomposition
  of emotional elements in speech,'' in \emph{2021 IEEE Spoken Language
  Technology Workshop (SLT)}.\hskip 1em plus 0.5em minus 0.4em\relax IEEE,
  2021, pp. 415--422.

\bibitem{zhou2021seen}
K.~Zhou, B.~Sisman, R.~Liu, and H.~Li, ``Seen and unseen emotional style
  transfer for voice conversion with a new emotional speech dataset,'' in
  \emph{ICASSP 2021-2021 IEEE International Conference on Acoustics, Speech and
  Signal Processing (ICASSP)}.\hskip 1em plus 0.5em minus 0.4em\relax IEEE,
  2021, pp. 920--924.

\bibitem{qian2019autovc}
K.~Qian, Y.~Zhang, S.~Chang, X.~Yang, and M.~Hasegawa-Johnson, ``Autovc:
  Zero-shot voice style transfer with only autoencoder loss,'' in
  \emph{International Conference on Machine Learning}.\hskip 1em plus 0.5em
  minus 0.4em\relax PMLR, 2019, pp. 5210--5219.

\bibitem{qian2020unsupervised}
K.~Qian, Y.~Zhang, S.~Chang, M.~Hasegawa-Johnson, and D.~Cox, ``Unsupervised
  speech decomposition via triple information bottleneck,'' in
  \emph{International Conference on Machine Learning}.\hskip 1em plus 0.5em
  minus 0.4em\relax PMLR, 2020, pp. 7836--7846.

\bibitem{chou2019one}
\BIBentryALTinterwordspacing
J.~chieh Chou and H.-Y. Lee, ``{One-Shot Voice Conversion by Separating Speaker
  and Content Representations with Instance Normalization},'' in \emph{Proc.
  Interspeech 2019}, 2019, pp. 664--668. [Online]. Available:
  \url{http://dx.doi.org/10.21437/Interspeech.2019-2663}
\BIBentrySTDinterwordspacing

\bibitem{chen2021again}
Y.-H. Chen, D.-Y. Wu, T.-H. Wu, and H.-y. Lee, ``Again-vc: A one-shot voice
  conversion using activation guidance and adaptive instance normalization,''
  in \emph{ICASSP 2021-2021 IEEE International Conference on Acoustics, Speech
  and Signal Processing (ICASSP)}.\hskip 1em plus 0.5em minus 0.4em\relax IEEE,
  2021, pp. 5954--5958.

\bibitem{wu2020vqvc+}
D.-Y. Wu, Y.-H. Chen, and H.-y. Lee, ``Vqvc+: One-shot voice conversion by
  vector quantization and u-net architecture,'' \emph{Proc. Interspeech 2020},
  pp. 4691--4695, 2020.

\bibitem{tjandra2019vqvae}
A.~Tjandra, B.~Sisman, M.~Zhang, S.~Sakti, H.~Li, and S.~Nakamura, ``Vqvae
  unsupervised unit discovery and multi-scale code2spec inverter for zerospeech
  challenge 2019,'' \emph{Proc. Interspeech 2019}, pp. 1118--1122, 2019.

\bibitem{wang2021vqmivc}
D.~Wang, L.~Deng, Y.~T. Yeung, X.~Chen, X.~Liu, and H.~Meng, ``{VQMIVC: Vector
  Quantization and Mutual Information-Based Unsupervised Speech Representation
  Disentanglement for One-Shot Voice Conversion},'' in \emph{Proc. Interspeech
  2021}, 2021, pp. 1344--1348.

\bibitem{zhou2021emotional}
K.~Zhou, B.~Sisman, R.~Liu, and H.~Li, ``Emotional voice conversion: Theory,
  databases and esd,'' \emph{Speech Communication}, vol. 137, pp. 1--18, 2022.

\bibitem{zhou2022emotion}
K.~Zhou, B.~Sisman, R.~Rana, B.~W. Schuller, and H.~Li, ``Emotion intensity and
  its control for emotional voice conversion,'' \emph{IEEE Transactions on
  Affective Computing}, 2022.

\bibitem{schuller2013computational}
B.~Schuller and A.~Batliner, \emph{Computational paralinguistics: emotion,
  affect and personality in speech and language processing}.\hskip 1em plus
  0.5em minus 0.4em\relax John Wiley \& Sons, 2013.

\bibitem{gan2022iqdubbing}
W.~Gan, B.~Wen, Y.~Yan, H.~Chen, Z.~Wang, H.~Du, L.~Xie, K.~Guo, and H.~Li,
  ``Iqdubbing: Prosody modeling based on discrete self-supervised speech
  representation for expressive voice conversion,'' \emph{arXiv preprint
  arXiv:2201.00269}, 2022.

\bibitem{yamamoto2020parallel}
R.~Yamamoto, E.~Song, and J.-M. Kim, ``Parallel wavegan: A fast waveform
  generation model based on generative adversarial networks with
  multi-resolution spectrogram,'' in \emph{ICASSP 2020-2020 IEEE International
  Conference on Acoustics, Speech and Signal Processing (ICASSP)}.\hskip 1em
  plus 0.5em minus 0.4em\relax IEEE, 2020, pp. 6199--6203.

\bibitem{van2008visualizing}
L.~Van~der Maaten and G.~Hinton, ``Visualizing data using t-sne.''
  \emph{Journal of machine learning research}, vol.~9, no.~11, 2008.

\bibitem{Lin2021S2VCAF}
J.~hao Lin, Y.~Y. Lin, C.~M. Chien, and H.~yi~Lee, ``S2vc: A framework for
  any-to-any voice conversion with self-supervised pretrained
  representations,'' in \emph{Interspeech}, 2021.

\end{thebibliography}

\end{document}